\documentclass[%
 reprint,
 amsmath,amssymb,
 aps,
]{revtex4-2}
\usepackage{amsmath, amssymb, amsthm}
\usepackage{enumitem}
\usepackage{physics}
\usepackage{bbm}
\usepackage[hidelinks]{hyperref}
\usepackage{tikz}
\usepackage{todonotes}
\usepackage{graphicx}
\usepackage{appendix}
\usepackage{dcolumn}
\usepackage{bm}

\usepackage{float}

\newcommand{\Ncal}{\mathcal{N}}

\newcommand{\Ecal}{\mathcal{E}}
\newcommand{\Rcal}{\mathcal{R}}
\newcommand{\Dcal}{\mathcal{D}}
\newcommand{\Vcal}{\mathcal{V}}

\newcommand{\C}{\mathbb{C}}

\newcommand{\ism}{\cong}
\newcommand{\N}{\mathbb{N}}

\newcommand{\adj}{\dagger}

\DeclareMathOperator{\Typ}{Typ}

\DeclareMathOperator{\Sym}{Sym}

\theoremstyle{plain}
\newtheorem{theorem}{Theorem}   
\newtheorem{lemma}[theorem]{Lemma}        
\newtheorem{corollary}[theorem]{Corollary} 
\newtheorem{proposition}[theorem]{Proposition}
\newtheorem{definition}[theorem]{Definition}

\begin{document}

\preprint{APS/123-QED}

\title{Optimal fidelity over zero-capacity erasure channels}

\author{Camilo Castro-Landolt$^1$}
\email{cecastro@uwaterloo.ca}
 
\author{Rabsan Galib Ahmed$^{2,3}$}%
 \email{rgahmed@uwaterloo.ca}

\author{Graeme Smith$^{2,3}$}
\email{graeme.smith@uwaterloo.ca}

\affiliation{$^1$Department of Physics and Astronomy, University of Waterloo, Ontario N2L 3G1, Canada.\\
$^2$Department of Applied Mathematics, University of Waterloo, Ontario N2L 3G1, Canada.\\
$^3$Institute for Quantum Computing, University of Waterloo, Ontario N2L 3G1, Canada.}

\date{\today}

\begin{abstract}
Antidegradable quantum channels have zero capacity for reliable quantum communication. However, they can nevertheless transmit quantum information with nontrivial fidelity. This raises the question: how well can quantum information be preserved through such channels? In this work, we exactly characterize the optimal asymptotic entanglement fidelity of quantum erasure channels throughout the antidegradable regime. Specifically, we show that it equals $\gamma+\frac{1-\gamma}{d}$ for transmission probabilities $0<\gamma\le \frac{1}{2}$ and logical dimension $d$. We use fractional extendibility to establish an upper bound on the entanglement fidelity achievable by arbitrary error-correcting protocols. Furthermore, we construct a family of permutation-invariant codes that asymptotically saturates this bound. 
\end{abstract}

\maketitle

\section{Introduction}
Quantum error correction (QEC) protects quantum information by distributing it across many physical systems \cite{PhysRevA.56.33,PhysRevLett.104.120501}. Increasing the encoding redundancy can significantly improve recovery; however, sufficiently severe noise imposes fundamental limitations on this approach. In particular, antidegradable channels \cite{Cubitt_2008}, for which the environment can simulate the receiver's output, cannot permit perfect recovery of quantum information, as this would violate the no-cloning theorem \cite{Wootters1982}. Consequently, such channels have zero quantum capacity, meaning they cannot reliably transmit quantum information at any positive asymptotic rate. Recent works have established an even stronger limitation, showing that, for any fixed positive transmission rate, the fidelity of any coding protocol decays exponentially with the number of channel uses \cite{cheng2026informationtransmissionquantumchannels,beigi2026strongconversesquantumchannel,kondra2026sharpquantumcapacitythresholds}. 

However, these limitations do not rule out nontrivial approximate recovery. Even when perfect recovery is impossible, increasing the encoding redundancy may still improve the fidelity of transmission. To understand the limits of this improvement, we consider a fixed-dimensional quantum system encoded across an increasing number of physical systems. In this setting, the transmission rate approaches zero, so the strong converse results do not determine the optimal achievable fidelity. This raises the question: how faithfully can a fixed quantum system be transmitted when arbitrarily many uses of a noisy channel are permitted? 

In this work, we investigate this question for the $d$-dimensional quantum erasure channel, defined as
\begin{align}
    \Ecal_\gamma(\rho)=\gamma\rho+(1-\gamma) \tr(\rho)\ket{e}\bra{e}.
\end{align}
This channel transmits an input state perfectly with probability $\gamma$ and, with probability $1-\gamma$, erases the input and replaces it with a flag $\ket{e}$ orthogonal to the input system \cite{Bennett1997}. For transmission probabilities $\gamma\le \frac{1}{2}$, it is antidegradable so perfect transmission is impossible.

To quantify transmission quality, we use entanglement fidelity, which measures the degree to which a channel preserves entanglement with a reference system \cite{PhysRevA.54.2614}. For a channel $\Ncal:A\to B$ with $A\ism B$, we define
\begin{align}\label{eq:entanglementfidelity}
    F_e(\Ncal):=\tr(\Phi_{RB}(I_R\otimes\Ncal)(\Phi_{RA})),
\end{align}
where $\Phi_{RA}$ is a normalized maximally entangled state between the input $A$ and an isomorphic reference system $R$. For a single use of $\Ecal_\gamma$, the optimal entanglement fidelity is achieved by applying a decoding operation $\Dcal$ that replaces the flag with a fixed state, giving $F_e(\Dcal\circ \Ecal_\gamma)=\gamma+\frac{1-\gamma}{d^2}$. This can be improved by codes that increase the number of channel uses. Previous work has established some limitations on such improvements: for symmetric erasure ($\gamma=\frac12$), transmission with entanglement fidelity greater than $\frac{1}{2}$ is possible only when the input dimension remains bounded, independently of the number of channel uses \cite{Beigi2016}. However, the optimal asymptotic entanglement fidelity for a fixed logical dimension and a corresponding characterization for $\gamma<\frac12$ were previously unknown. 

For a fixed input dimension $d$, we define the optimal entanglement fidelity achievable over $n$ uses of $\mathcal{E}_\gamma$ by
\begin{align}
F_n^\star(\gamma,d):=\sup_{\Vcal,\Dcal}F_e(\Dcal\circ \Ecal_\gamma^{\otimes n}\circ \Vcal),
\end{align}
where the supremum is taken over all completely positive trace-preserving (CPTP) encoding and decoding operations $\Vcal$ and $\Dcal$. Our main result exactly characterizes the asymptotic optimal entanglement fidelity of the quantum erasure channel throughout the antidegradable regime. 
\begin{theorem}[Main result]\label{thm:main}
    For all $d\ge 2$, $\gamma\in\left(0,\frac12\right]$ and $n\in \N$, the optimal entanglement fidelity satisfies
    \begin{align}
        F_n^\star(\gamma,d)\le\gamma+\frac{1-\gamma}{d}.
    \end{align}
    Moreover, this bound is asymptotically achievable:
    \begin{align}
        \lim_{n\to\infty }F_n^\star(\gamma,d)=\gamma+\frac{1-\gamma}{d}.
    \end{align}
\end{theorem}
 The bound is established by making use of fractional extendibility, which was recently introduced for Gaussian states in \cite{Ahmed2026NoGoTF} and subsequently developed for general quantum states in \cite{Ahmed2026}. We derive a bound on the singlet fraction of fractionally extendible states and apply it to arbitrary encoding and decoding protocols for the erasure channel.
    
    To establish achievability, we turn to permutation-invariant (PI) codes, a well established approach to QEC, with early constructions by Ruskai and Pollatsek~\cite{Ruskai2000,PollatsekRuskai2004}. Later work has focused on developing more general families of PI codes and investigated their structural properties~\cite{Ouyang2014,OuyangFitzsimons2016,OuyangBrennen2026}, as well as their applications to erasure and deletion error correction~\cite{Ouyang2021,shibayama2021permutation}.
    We construct a family of PI codes using qudit Dicke states that achieves the optimal fidelity by balancing distinguishability of logical basis states against preservation of coherence between their superpositions.

\section{Preliminaries}
\subsection{Fractional extendibility}
The $k$-extendibility of a state characterizes the extent to which quantum correlations with a reference can be shared \cite{Doherty2004,Brand_o_2012,Li_2018,Kaur_2019}. A bipartite state $\rho_{RB}$ is said to be $k$-extendible with respect to the subsystem $B$, for some integer $k\ge 1$, if there exists a state $\rho_{RB_1B_2\dots B_k}$ such that $\rho_{RB_i}=\rho_{RB}$ for every $1\le i\le k$. Fractional extendibility generalizes this notion by allowing the correlations of a bipartite state to be recovered from subsets of a multipartite extension.

For a positive integer $n$ we define $[n]:=\{1,\dots,n\}$ and use $S\subseteq_k[n]$ to denote a subset of size $k\le n$.
\begin{definition}[Fractional extendibility]\label{def:fractional extendibility}
    Let $p\ge q\ge 1$ be integers. A bipartite state $\rho_{RB}$ is $(p,q)$-extendible if there exists a state $\sigma_{RB_1\cdots B_{p+q}}$ such that for every subset $S\subseteq_q[p+q]$, there exists a channel $\Rcal_S:B_S\to D_S$, where $B_S=\bigotimes_{i\in S}B_i$ and $D_S\ism B$, satisfying $(I_R\otimes \Rcal_S)(\sigma_{RB_S})=\rho_{RD_S}$.
\end{definition}
Operationally, $(p,q)$-extendibility means that the correlations between $R$ and $B$ can be recovered from any $q$ of the $p+q$ extension systems. In particular, since $p\ge q$, this allows two disjoint subsets to simultaneously reproduce the same correlations with $R$. This places constraints on the degree of entanglement of $\rho_{RB}$.

We now give an outline of the properties of fractional extendibility proven in \cite{Ahmed2026}, as well as the tools developed there that are useful to apply the idea to our problem.

Fractional extendibility is preserved under tensor products. That is, if $\rho_{RB}$ is $(p,q)$-extendible, then so too is $\rho_{RB}^{\otimes n}$ for every $n\ge 1$. Additionally, extendibility is preserved under local operations on $R$ or $B$.

We call a quantum channel $\Ncal$ $(p,q)$-extendible when its Choi state $(I_R\otimes \Ncal)(\Phi_{RA})$ is $(p,q)$-extendible. The corresponding channel properties imply that bounds applying to $(p,q)$-extendible states can be used to constrain arbitrary encoding and decoding protocols for $(p,q)$-extendible channels. In particular, an erasure channel with transmission probability $\frac{q}{p+q}$ is $(p,q)$-extendible.

To obtain quantitative bounds on recoverable entanglement, we make use of graph-theoretic constructions. For a graph $G=(V,E)$, a $\mu$-orthonormal representation associates a unit vector $\ket{v_i}$ to each $i\in V$, such that, whenever $\{i,j\}\in E$, $\abs{\bra{v_i}\ket{v_j}}\le \mu$. The $\mu$-approximate Lov\'{a}sz theta function is defined by
\begin{align}\label{def:approxlovasztheta}
    \vartheta_\mu(G):=\max_{\ket{\psi},\{\ket{v_i}\}}\sum_{i\in V}\abs{\bra{\psi}\ket{v_i}}^2,
\end{align}
where the maximum is taken over all $\mu$-orthonormal representations $\{\ket{v_i}\}$ and unit vectors $\ket{\psi}$. It provides an upper bound on the total squared overlap that the vectors of a $\mu$-orthonormal representation can have with a common reference vector.

The overlap constraints associated with disjoint recovery subsets of a $(p,q)$-extension can be described using Kneser graphs. The Kneser graph $KG_{n,k}$ has vertices corresponding to the $k$-subsets of $[n]$ and edges between vertices whose subsets are disjoint. For $0<\mu<1$ and $n\ge 2k$, it was shown in~\cite{Ahmed2026} that
\begin{align}\label{eq:explicitlovasz}
    \vartheta_\mu\left(KG_{n,k}\right)=\binom{n-1}{k-1}\left(1+\mu\frac{n-k}{k}\right).
\end{align}
\subsection{Permutation-invariant codes}

In this work we consider PI encodings that map a $d$-dimensional logical input into the symmetric subspace $\Sym^n(\C^d)$ of $n$ physical qudits. The resulting encoded states are invariant under permutations of the physical subsystems which simplifies the analysis of erasure errors. In particular, when $\Ecal_\gamma$ acts independently on each physical system, erasure patterns with the same number $j$ of erased systems yield equivalent reduced states, up to relabelling \cite{Ouyang2021}. We may therefore, without loss of generality, choose the final $j$ systems to be the ones erased, which also allows a decoding operation to be conditioned on $j$. 

The symmetric subspace admits an orthonormal basis of qudit Dicke states \cite{Dicke1954,Nepomechie2024}, which are equal superpositions of computational basis states with fixed occupation numbers. Let $\vec{r}=(r_0,\cdots,r_{d-1})\in \N^d_0$ with $\sum_{a=0}^{d-1}r_a=n$, where $\N_0^d$ is the set of vectors of length $d$ with non-negative integer entries and $r_a$ denotes the number of occurrences of the basis label $a$. The corresponding qudit Dicke state is defined by 
\begin{align}
    \ket{D^n(\vec r)}:=\sqrt{\frac{\prod_{a=0}^{d-1}r_a!}{n!}}\sum_{P}P(\ket{0}^{\otimes r_0}\cdots\ket{d-1}^{\otimes r_{d-1}}),
\end{align}
where the sum is over all distinct permutations $P$ of the tensor factors. 

Dicke states with distinct occupation vectors are orthogonal. We denote the standard basis vector with a single non-zero entry in position $a$ by $e_a\in \N_0^d$. 

\section{main results}
\subsection{Upper bound on the optimal entanglement fidelity}\label{sct:bound}

\begin{theorem}\label{thm:bound}
Let $\rho_{RB}$ be a $(p,q)$-extendible state with $R\ism B\ism \C^d$. Then
\begin{align}
    \tr(\Phi_{RB}\rho_{RB})\le \frac{q}{p+q}+\frac{p}{d(p+q)}
\end{align}
\end{theorem}
\begin{proof}
    Let $\rho_{RB}$ be $(p,q)$-extendible. Then, there exists a state $\sigma_{RB_1\cdots B_{p+q}}$ such that, for every $S\subseteq_q[p+q]$, there is a recovery channel $\Rcal_S:B_S\to D_S$ with $D_S\ism B\ism \C^d$, such that $(I_R\otimes \Rcal_S)(\sigma_{RB_S})=\rho_{RD_S}$.
    Purify $\sigma$ to a state $\ket{\sigma}_{RB_{[p+q]}E}$. For each $S\subseteq_q[p+q]$, prepare a fixed pure state $\ket{0}_{K_S}$ in a large enough ancillary system $K_S$ so that the Stinespring dilation $U_S:B_SK_S\rightarrow D_SE_S$ of $\Rcal_S$ can be taken to be unitary. Hence, $\Rcal_S(X)=\tr_{E_S}(U_S(X\otimes \ket{0}\bra{0}_{K_S})U^\adj_S)$. We regard $U_S$ as implicitly tensored with the identity on all remaining systems.\\ Define $\ket{\tilde\sigma}:=\ket{\sigma}_{RB_{[p+q]}E}\otimes\bigotimes _{S\subseteq_q[p+q]}\ket{0}_{K_S}.$
    For each $S$, define the maximally entangled projector $\Pi_S:=\Phi_{RD_S}\otimes I$ and $M_S:=U_S^\adj \Pi_S U_S$. Since $\Pi_S$ is an orthogonal projector and $U_S$ is unitary, $M_S$ is also an orthogonal projector. Moreover,
    \begin{align}
\bra{\tilde\sigma}M_S\ket{\tilde\sigma}&=\tr[\Phi_{RD_S}(I_R\otimes \Rcal_S)(\sigma_{RB_S})]\nonumber\\&=\tr(\Phi_{RD_S}\rho_{RD_S})\nonumber\\&=\tr(\Phi_{RB}\rho_{RB}).
    \end{align}
    For ease, let $F=\tr(\Phi_{RB}\rho_{RB})$. If $F=0$ the claim follows trivially, so assume $F>0$ and define $\ket{v_S}:=\frac{M_S\ket{\tilde\sigma}}{\sqrt{F}}.$
    Since $M_S=M_S^\adj=M_S^2$, $\bra{v_S}\ket{v_S}=1$ and $\abs{\bra{\tilde\sigma}\ket{v_S}}^2=F$. Now let $S,T\subseteq_q[p+q]$ with $S\cap T=\emptyset$. Since their corresponding dilations act on disjoint subsystem, they commute with each other, and each commutes with the maximally entangled projector associated with the other subset. Thus
    \begin{align}
        M_SM_T=(U_SU_T)^\adj \Pi_S\Pi_T(U_SU_T).\nonumber
    \end{align}
    Applying Lemma~\ref{lem:proj} proven in Appendix~\ref{app:overlap-lemma}, yields
    \begin{align}
        \norm{M_SM_T}_{\infty}=\norm{\Pi_S\Pi_T}_{\infty}=\frac{1}{d}.
    \end{align}
    Where $\norm{X}_\infty:=\sup_{\norm{\ket{\psi}}=1}\norm{X\ket{\psi}}$ denotes the operator norm.\\
    It follows that 
    \begin{align}
        \abs{\bra{v_S}\ket{v_T}}&=\frac{\abs{\bra{\tilde\sigma}M_SM_T\ket{\tilde\sigma}}}{F}\nonumber\\
        &=\frac{\abs{\bra{M_S\tilde\sigma}M_SM_T\ket{M_T\tilde\sigma}}}{F}\nonumber\\&\le\frac{\norm{M_S\ket{\tilde\sigma}}\norm{M_T\ket{\tilde\sigma}}\norm{M_SM_T}_{\infty}}{F}\nonumber\\&=\frac{1}{d},
    \end{align}
    where the inequality follows from the Cauchy-Schwarz inequality and the definition of the operator norm.
    
    Thus, the set $\{\ket{v_S}:S\subseteq_q[p+q]\}$ is a $\frac{1}{d}$-orthonormal representation of the Kneser graph $KG_{p+q,q}$. Since $p\ge q$ implies $p+q\ge 2q$ and clearly $0\le \frac{1}{d}\le 1$, we can make use of Equation~\eqref{eq:explicitlovasz}.
    \begin{align}
        \binom{p+q}{q}F&=\sum_{S\subseteq_q[p+q]}\abs{\bra{\tilde\sigma}\ket{v_S}}^2\nonumber\\&\le\vartheta_{\frac1d}(KG_{p+q,q})\nonumber\\&=\binom{p+q-1}{q-1}\left(1+\frac{p}{qd}\right),
    \end{align}
    and thus,
    \begin{align}
        \tr(\Phi_{RB}\rho_{RB})&\le \frac{\binom{p+q-1}{q-1}}{\binom{p+q}{q}}\left(1+\frac{p}{qd}\right)\nonumber\\&=\frac{q}{p+q}+\frac{p}{d(p+q)}.\nonumber
    \end{align}
\end{proof}
\noindent The upper bound in Theorem~\ref{thm:main} is established by the following corollary.
\begin{corollary}\label{cor:bound}
    For all $d\ge 2$, $n\in \N$ and $\gamma\in \left(0,\frac{1}{2}\right]$,
    \begin{align}
        F_n^\star(\gamma,d)\le \gamma+\frac{1-\gamma}{d}.
    \end{align}
\end{corollary}
\begin{proof}
    For rational $\gamma=\frac{q}{p+q}$ where $p,q\in \N$ and $p\ge q$, the result follows immediately from Theorem~\ref{thm:bound}. As shown in \cite{Ahmed2026}, the Choi state $(I_R\otimes\Ecal_{\gamma})\Phi_{RA}$ is $(p,q)$-extendible and, for any encoding $\Vcal$ and decoding $\Dcal$, the state $\rho_{RB}:=(I_R\otimes\Dcal\circ \Ecal_\gamma^{\otimes n}\circ \Vcal)(\Phi_{RA}) $ is $(p,q)$-extendible as well. Thus, the upper bound on the singlet fraction of $\rho_{RB}$, as found in Theorem~\ref{thm:bound}, provides an upper bound on the achievable entanglement fidelity over arbitrarily many uses of $\Ecal_\gamma$:
    \begin{align}   F_n^\star(\gamma,d)&=\sup_{\rho_{RB}}\Tr(\Phi_{RB}\rho_{RB})\nonumber\\&\le \frac{q}{p+q}+\frac{p}{d(p+q)}\nonumber\\&=\gamma+\frac{1-\gamma}{d}\nonumber.
    \end{align}
    Now let $\gamma$ be irrational and choose a sequence of rational numbers $\gamma_m$ such that $\gamma< \gamma_m\le \frac{1}{2}$ and $\gamma_m\to \gamma$. Letting $\delta_m:=\frac{\gamma}{\gamma_m}$ we have $\Ecal_\gamma=\Ecal_{\delta_m}\circ \Ecal_{\gamma_m}$. The $\Ecal_{\delta_m}$ operation can be absorbed into a decoding operation. Hence, any entanglement fidelity achievable over $n$ uses of $\Ecal_\gamma$ is also achievable over $n$ uses of $\Ecal_{\gamma_m}$. Thus 
    \begin{align}
        F_n^\star(\gamma,d)\le F^\star_n(\gamma_m,d)\le\gamma_m+\frac{1-\gamma_m}{d}.
    \end{align}
    Taking $m\to \infty$ proves the claim.
\end{proof}
\subsection{Code achieving the upper bound}
We now construct an explicit family of PI codes that asymptotically saturates the bound in Corollary~\ref{cor:bound}. Throughout this section we fix $n\ge 3$ and $t\in [0,1]$. For $a\in\{0,\dots,d-1\}$, let
\begin{align}\label{eq:enc}
V_{n,t}\ket{a}&:=\sqrt{t}\ket{D^n(ne_a)}\nonumber\\&\quad+\sqrt{\frac{1-t}{d-1}}\sum_{\substack{k=0\\k\ne a}}^{d-1}\ket{D^n(e_a+(n-1)e_k)}.
\end{align}
 We define our encoding map by $\Vcal_{n,t}(\rho):=V_{n,t}\rho V_{n,t}^\adj$.
 
The encoding is constructed to balance the competing requirements of preserving distinguishability between logical basis states and maintaining coherence between their superpositions.

The first component consists of repetition states $\ket{D^n(ne_a)}=\ket{a}^{\otimes n}$ which redundantly encode the logical basis label $a$, so that it can be recovered from any transmitted subset. However, since every physical subsystem carries the logical basis label, erasing even a single subsystem destroys coherence between distinct repetition states. 

The second component consists of Dicke states with a single excitation $a$ against a background of $n-1$ copies of another basis label $k$. Unlike the repetition states, these configurations can preserve coherence between different logical labels when the erased subsystems contain only their common background. For a fixed number $j$ of erasures, the excitation survives with probability $1-\frac{j}{n}$, allowing the logical label to be recovered from the surviving systems. 

\begin{proposition}\label{prop:isometry}
For $n\geq3$, $V_{n,t}$ is an isometry.
\end{proposition}

\begin{proof}
For fixed $a$, all Dicke states appearing in Eq.~\eqref{eq:enc} have distinct occupation vectors, and are thus mutually orthogonal. Hence
\begin{align}
\|V_{n,t}\ket a\|^2=t+\frac{1-t}{d-1}(d-1)=1.
\end{align}
For $a\neq b$, no occupation vector appearing in $V_{n,t}\ket a$ appears in $V_{n,t}\ket b$ when $n\geq3$. Therefore $\bra aV_{n,t}^\adj V_{n,t}\ket b=0$ and thus $V_{n,t}^\adj V_{n,t}=I_d$.
\end{proof}

We condition the decoder on a fixed number $j$ of erasures. For each $i\in \{0,\dots,d-1\}$, define
\begin{align}\label{eq:decoderprojectors}
    D_{n,j}^{(i)}:=\sum_{a=0}^{d-1}\ket{a}\bra{D^{n-j}(e_a+(n-j-1)e_i)}.
\end{align}
Let $ P_{n,j}:=\sum_{i=0}^{d-1}\left(D_{n,j}^{(i)}\right)^\adj D_{n,j}^{(i)}$ and $Q_{n,j}:=I-P_{n,j}$, and define
\begin{align}\label{eq:dec}
    \Dcal_{n,j}(\rho):=\sum_{i=0}^{d-1}D_{n,j}^{(i)}\rho \left(D_{n,j}^{(i)}\right)^\adj+\tr(Q_{n,j}\rho)\frac{I_d}{d}.
\end{align}
For each background label $i$, the operator $D_{n,j}^{(i)}$ maps the corresponding Dicke states to their logical basis labels. In particular, when $i=a$, this recovers the repetition states $\ket{a}^{\otimes (n-j)}$.
Each $D_{n,j}^{(i)}$ maps the $d$ states associated with the fixed background $i$ coherently into the logical space so that off diagonal terms between the configurations are preserved. The $Q_{n,j}$ term ensures that the decoder is trace preserving by associating a maximally mixed state to configurations outside of the designated decoding subspace.
\begin{proposition}\label{prop:dec-cptp}
If $n-j\geq3$, then $\Dcal_{n,j}$ is CPTP.
\end{proposition}

\begin{proof}
The first term in Eq.~\eqref{eq:dec} is completely positive \cite{wilde2017quantum}. For $n-j\geq3$, the occupation vectors $e_a+(n-j-1)e_i$ are all distinct. Thus, the Dicke states appearing in Eq.~\eqref{eq:decoderprojectors} form an orthonormal set. It follows that $P_{n,j}$ is an orthogonal projector, so $Q_{n,j}\succeq0$.
 Thus the map $\rho\mapsto\tr(Q_{n,j}\rho)\frac{I_d}{d}$ is also completely positive. Finally, $\tr(\Dcal_{n,j}(\rho))=\tr(P_{n,j}\rho)+\tr(Q_{n,j}\rho)=\tr(\rho),$ so $\Dcal_{n,j}$ is trace preserving.
\end{proof}
For $3\le j\le n-3$, define $\Ncal_{n,j,t}:=\Dcal_{n,j}\circ \tr_j\circ \Vcal_{n,t}$, where $\tr_j$ traces out the final $j$ systems once the erasure pattern has been identified via the flag $\ket{e}$.
\begin{proposition}\label{prop:logicalchannel}
    For $3\le j\le n-3$, the conditional logical channel acts on the computational-basis matrix elements as follows. For the diagonal terms,
    \begin{align}\label{eq:logicaldiag}
        \Ncal_{n,j,t}(\ket{a}\bra{a})&=\left[t+(1-t)\left(1-\frac{j}{n}\right)\right]\ket{a}\bra{a}\nonumber\\&\quad+\frac{1-t}{d-1}\frac{j}{n}\sum_{\substack{k=0\\k\ne a}}^{d-1}\ket{k}\bra{k},
    \end{align}
    and for $a\ne b$,
    \begin{align}\label{eq:logicaloffdiag}
        \Ncal_{n,j,t}(\ket{a}\bra{b})&=\Bigg[2\sqrt{\frac{t(1-t)}{d-1}\left(1-\frac{j}{n}\right)}\nonumber\\&+\frac{(1-t)(d-2)}{d-1}\left(1-\frac{j}{n}\right)\Bigg]\ket{a}\bra{b}.
    \end{align}
\end{proposition}
For the sake of readability, the full calculation is left to appendix~\ref{appendixA}.

The diagonal action describes the recovery of logical basis information. When at least one physical subsystem survives erasure, the repetition component contributes $t$ to the probability of recovering $\ket{a}$. The  single-excitation component contributes $(1-t)\left(1-\frac{j}{n}\right)$, corresponding to the probability that the excitation survives erasure. When the excitation is erased, the decoder recovers its background label.

The off diagonal action describes the preservation of coherence between distinct logical basis states. The first contribution arises from interference between repetition and single-excitation components with a common background label. For $a\ne b$, these correspond to pairings between the repetition state encoding $a$ and the excitation state encoding $b$ with  background $a$, and vice versa. The second contribution arises from pairs of single excitation states encoding distinct logical labels $a$ and $b$ against a common background. 

The entanglement fidelity of the logical channel captures the contributions from both~\eqref{eq:logicaldiag} 
and~\eqref{eq:logicaloffdiag} allowing us to determine the optimal encoding weight $t$.
\begin{proposition}\label{prop:conditionalfe}
    For $3\le j\le n-3$,
    \begin{align}\label{eq:conditionalfe}
        F_{e,j,n}(t)&:=F_e(\Ncal_{n,j,t})\nonumber\\&=\frac{1}{d}\Bigg[\sqrt{t}+\sqrt{(d-1)(1-t)\left(1-\frac{j}{n}\right)}\Bigg]^2.
    \end{align}
\end{proposition}
\begin{proof}
    It is easy to see from Equation~\eqref{eq:entanglementfidelity} that, for a $d$-dimensional input,
    \begin{align}
    F_e(\Ncal_{n,j,t})=\frac{1}{d^2}\sum_{a,b=0}^{d-1}\bra{a}\Ncal_{n,j,t}(\ket{a}\bra{b})\ket{b}.
    \end{align}
    The $d$ diagonal matrix elements each contribute $\bra{a}\Ncal_{n,j,t}(\ket{a}\bra{a})\ket{a}=t+(1-t)\left(1-\frac{j}{n}\right)$ and the $d(d-1)$ off-diagonal matrix elements each contribute $\bra{a}\Ncal_{n,j,t}(\ket{a}\bra{b})\ket{b}
        =2\sqrt{\frac{t(1-t)}{d-1}\left(1-\frac{j}{n}\right)}+\frac{(1-t)(d-2)}{d-1}\left(1-\frac{j}{n}\right)$. Combining these yields
    \begin{align}
        F_{e,j,n}(t)=\frac{1}{d}\left[\sqrt{t}+\sqrt{(d-1)(1-t)\left(1-\frac{j}{n}\right)}\right]^2.
    \end{align}
\end{proof}
The conditional entanglement fidelity depends on the erasure number $j$ only through the fraction $1-\frac{j}{n}$ of surviving physical systems. To express this dependence, for $x\in [0,1]$, define
\begin{align}
    f_t(x):=\frac{1}{d}\left[\sqrt{t}+\sqrt{(d-1)(1-t)x}\right]^2.
\end{align}
Then, for $3\le j\le n-3$, $F_{e,j,n}(t)=f_t\left(1-\frac{j}{n}\right)$.

To determine the appropriate encoding parameter for the full channel, we must account for the distribution of erasure outcomes. Fix $\gamma\in\left(0,\frac{1}{2}\right]$ and write the total number of erasures $J_n=\sum_{i=1}^{n}X_i$, where the $X_i$'s are independent Bernoulli random variables that indicate whether the $i$-th system is erased:
\begin{align}
    X_i:=\begin{cases}
        1&\text{with probability $1-\gamma$}\\
        0&\text{with probability $\gamma$}
    \end{cases}.
\end{align}
For $\delta>0$ define the $\delta$-typical erasure number set for $n$ channel uses by
\begin{align}\label{eq:typ}
    \Typ_\delta^n:=\left\{j\in \{0,\dots,n\}:\abs{\frac{j}{n}-(1-\gamma)}\le \delta\right\}
\end{align}
Hoeffding's inequality \cite{Hoeffding1963} then gives 
\begin{align}\label{eq:typicalbound}
    \sum_{j\in \Typ_\delta^n}\Pr[J_n=j]\ge 1-2\exp(-2n\delta^2).
\end{align}

It follows that, as $n$ increases the fraction of surviving physical systems becomes concentrated around $\gamma$. This motivates choosing the encoding parameter $t$ to maximize $f_t(\gamma)$. Applying the Cauchy-Schwarz inequality to the vectors $(\sqrt{t},\sqrt{1-t})$ and $(1,\sqrt{(d-1)\gamma})$, we find that $f_t(\gamma)$ is maximized at $t^\star(d,\gamma):=\frac{1}{1+(d-1)\gamma}$ with maximum $f_{t^\star}(\gamma)=\gamma+\frac{1-\gamma}{d}$.
This coincides with the upper bound established in Corollary~\ref{cor:bound}. It now only remains to show that our code achieves this value asymptotically.

\begin{theorem}[Asymptotic achievability]\label{thm:achievability} Fix $d\ge 2$ and $\gamma\in \left(0,\frac{1}{2}\right]$, and let $t^\star=\frac{1}{1+(d-1)\gamma}$. Then there exists a sequence of decoders $\{\Dcal_n\}_{n\ge 3}$ for the encodings $\Vcal_{n,t^\star}$ such that
\begin{align}
  \lim_{n\to\infty}F_e(\Dcal_n\circ \Ecal_\gamma^{\otimes n}\circ \Vcal_{n,t^\star})= \gamma+\frac{1-\gamma}{d}
\end{align}
\end{theorem}
\begin{proof}
    The erasure flags allow the receiver to determine $j$ without disturbing the surviving systems. For $3\le j\le n-3$, apply the decoder $\Dcal_{n,j}$ constructed above. For the remaining erasure numbers choose arbitrary CPTP decoding maps with the appropriate input and output spaces. When combined, these conditional maps define a CPTP decoder $\Dcal_n$ for the full erasure channel.  Let $\delta_n=\sqrt{\frac{\ln n}{n}}$.
    Since $\gamma\le \frac{1}{2}$ is fixed and $\delta_n\to 0$, for sufficiently large $n$, we have $\gamma-\delta_n\ge \frac3n$ and $1-\gamma-\delta_n\ge \frac{3}{n}$. Thus, every $j\in \Typ_{\delta_n}^n$ satisfies $3\le j\le n-3$, so the entanglement fidelity expression in Proposition~\ref{prop:conditionalfe} applies throughout the typical set. By linearity of entanglement fidelity
    \begin{align}
    F_e(\Dcal_n\circ\Ecal_\gamma^{\otimes n}\circ \Vcal_{n,t^\star})&\ge \sum_{j\in\Typ^n_{\delta_n}}\Pr[J_n=j]f_{t^\star}\left(1-\frac{j}{n}\right)\nonumber\\&\ge \left(1-\frac{2}{n^2}\right)f_{t^\star}(\gamma-\delta_n)\nonumber.
    \end{align}
    Here, the second inequality follows from the monotonicity of $f_{t^\star}(x)$, since $1-\frac{j}{n}\ge \gamma-\delta_n$ throughout the typical set, and from applying Equation~\eqref{eq:typicalbound}. Thus, since $\delta_n\to 0$ and $f_{t^\star}$ is continuous,
    \begin{align}
    \liminf_{n\to\infty}F_e(\Dcal_n\circ\Ecal_\gamma^{\otimes n}\circ \Vcal_{n,t^\star})\ge f_{t^\star}(\gamma)=\gamma+\frac{1-\gamma}{d}.
    \end{align} 
    On the other hand, Corollary~\ref{cor:bound} gives, for every $n$,
    \begin{align}
        F_e(\Dcal_n\circ \Ecal_\gamma^{\otimes n}\circ \Vcal_{n,t^\star})\le F_n^\star (\gamma,d)\le \gamma+\frac{1-\gamma}{d}.
    \end{align}
    Combining the two bounds proves the claim.
\end{proof}
Theorem~\ref{thm:achievability}, together with Corollary~\ref{cor:bound}, establishes the main result in Theorem~\ref{thm:main}.

Having established the asymptotic optimality of our code, we now have an intuitive way to understand how the optimized parameter $t^\star(d,\gamma)=\frac{1}{1+(d-1)\gamma}$  distributes weight between the repetition and single-excitation components of the encoding. 

\begin{figure}[!t]
    \centering
    \includegraphics[width=\columnwidth]{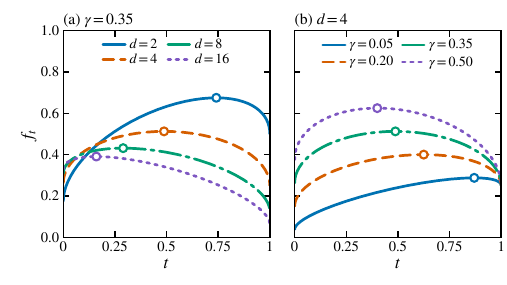}
    \caption{Asymptotic entanglement fidelity $f_t$ of the conditional logical channel as a function of the encoding parameter $t$. (a) Fidelity for different logical dimensions $d$ and fixed transmission probability $\gamma=0.35$. (b) Fidelity for different transmission probabilities $\gamma$ at fixed logical dimension $d=4$. Open circles on each curve indicate the entanglement fidelity achieved by the optimal encoding parameter $t^\star(d,\gamma)=\frac{1}{1+(d-1)\gamma}$.}
    \label{fig:fidelity}
\end{figure}

For fixed transmission probability $\gamma$, the relative importance of off-diagonal contributions to the entanglement fidelity grows with logical dimension $d$, making it increasingly favourable to preserve coherence over logical basis information. The optimal encoding thus assigns greater weight to the single-excitation components, shifting the optimum towards smaller values of $t$, as illustrated in Fig.~\ref{fig:fidelity} (a).

For fixed $d$, lower transmission probabilities make it less likely that an excitation survives erasure, reducing the effectiveness of the single-excitation components in preserving coherence. The optimal encoding therefore shifts weight towards the repetition states, whose preservation of logical basis labels is more robust to erasure errors. Conversely, as $\gamma$ increases, excitations become more likely to survive, and the optimal encoding assigns greater weight to the single-excitation components, as seen in Fig~\ref{fig:fidelity} (b).  

\section{Discussion}
We have established a complete characterization of the asymptotic optimal entanglement fidelity of the $d$-dimensional quantum erasure channel in the antidegradable regime. Using fractional extendibility, we derived an upper bound on the singlet fraction of $(p,q)$-extendible states, which in turn gives a bound on the entanglement fidelity achievable by arbitrary QEC protocols for the erasure channel. We showed this bound is asymptotically tight by constructing an explicit family of PI codes using qudit Dicke states. 

Beyond erasure, the singlet fraction bound in Theorem~\ref{thm:bound} provides a tool for studying approximate QEC for other classes of $(p,q)$-extendible channels.
In particular, our result can be easily extended to apply to antidegradable qubit damping (AD) channels \cite{Nielsen2010,PhysRevA.56.2567}. As discussed in \cite{Ahmed2026}, an AD channel with damping probability $\eta=\frac{p}{p+q}$ admits a symmetric $(p,q)$-extension. Our bound therefore implies its optimal entanglement fidelity is at most $1-\frac{\eta}{2}$. Conversely, using a dual-rail encoding \cite{Duan2010MultiErrorCorrecting,Kubica2023}, one can simulate a qubit erasure channel with two uses of an AD channel. Combining this simulation with our erasure channel code gives a corresponding asymptotically tight characterization for antidegradable qubit AD channels. 

\begin{acknowledgments}
In June of 2026, AI tools were used to assist with numerical optimization intended to search for promising error correcting codes during the initial exploration of the special case of a qubit erasure channel with 50\% transmission probability. All analytical calculations, derivations and proofs were carried out independently by the authors without AI assistance.
This work is supported under NSERC-NSF alliance grant ALLRP-586858-2023 and NSERC Discovery grant
RGPIN-2025-02094. RGA acknowledges the support of the Institute for Quantum Computing and Mike and Ophelia Lazaridis Graduate Fellowship. After the manuscript was fully prepared, the authors became aware of a recent work by Gamal et.al.~\cite{gamal2026forwardbitmakeserror} which provides a similar fidelity bound for the anti-degradable qubit amplitude damping channel.
\end{acknowledgments}

\appendix
\section{Full logical channel calculation}\label{appendixA}
Here we give the calculation for Proposition~\ref{prop:logicalchannel}. Fix $3\le j\le n-3$ and, for compactness, let $m=n-j$, $x:=\frac{m}{n}=1-\frac{j}{n}$, $y:=\frac{j}{n}$, $r:=\frac{1-t}{d-1}$ and define $\ket{A_a^n}:=\ket{D^n(ne_a)}$ and $\ket{B_{a,k}^n}:=\ket{D^n(e_a+(n-1)e_k)}$. The action of the encoding channel on an arbitrary matrix element with $a,b\in\{0,\dots,d-1\}$ is given by
\begin{align}\label{eq:encodingaction}
    &\Vcal_{n,t}(\ket{a}\bra{b})=t\ket{A_a^n}\bra{A_b^n}+\sqrt{rt}\sum_{\substack{k=0\\k\ne a}}^{d-1}\ket{B_{a,k}^n}\bra{A_b^n}\nonumber\\&+\sqrt{rt}\sum_{\substack{\ell=0\\\ell\ne b}}^{d-1}\ket{A_a^n}\bra{B_{b,\ell}^n} +r\sum_{\substack{k=0\\k\ne a}}^{d-1}\sum_{\substack{\ell=0\\\ell\ne b}}^{d-1}\ket{B_{a,k}^n}\bra{B_{b,\ell}^n}.
\end{align}
From here we trace out the final $j$ systems term by term. To this end, the two relevant Dicke states can be decomposed as\cite{Nepomechie2024}
\begin{align}\label{Dickedecomps}
&\ket{A_a^n}=\ket{A_a^m}\ket{A_a^j},\\&\ket{B_{a,k}^n}=\sqrt{x}\ket{B_{a,k}^m}\ket{A_k^j}+\sqrt{y}\ket{A_k^m}\ket{B_{a,k}^j}.
\end{align}
For the first term
\begin{align}\label{traceterm1}
\tr_j(\ket{A_a^n}\bra{A_b^n})=\delta_{ab}\ket{A_a^m}\bra{A_a^m}.
\end{align}
For the second term, orthogonality of the erased occupation vectors gives, since $j\ge 3$,
\begin{align}\label{traceterm2}
\tr_j(\ket{B_{a,k}^n}\bra{A_b^n})=\sqrt{x}\delta_{kb}\ket{B_{a,k}^m}\bra{A_b^m}.
\end{align}
The third trace term is similarly
\begin{align}\label{traceterm3}
\tr_j(\ket{A_a^n}\bra{B_{b,\ell}^n})=\sqrt{x}\delta_{\ell a}\ket{A_a^m}\bra{B_{b,\ell}^m}.
\end{align}
Finally for the fourth term, the cross terms between the two branches of the Dicke decomposition vanish after the partial trace. The remaining overlaps satisfy $\bra{A_\ell^j}\ket{A_k^j}=\delta_{\ell k}$ and $\bra{B_{b,\ell}^j}\ket{B_{a,k}^j}=\delta_{ab}\delta_{\ell k}$. Hence
\begin{align}\label{traceterm4}
    &\tr_j(\sum_{\substack{k=0\\k\ne a}}^{d-1}\sum_{\substack{\ell=0\\\ell\ne b}}^{d-1}\ket{B_{a,k}^n}\bra{B_{b,\ell}^n})\nonumber\\
    &=x\sum_{\substack{k=0\\k\ne a,b}}^{d-1}\ket{B_{a,k}^m}\bra{B_{b,k}^m}+y\delta_{ab}\sum_{\substack{k=0\\k\ne a}}^{d-1}\ket{A_k^m}\bra{A_k^m}.
\end{align}
The decoder acts on the relevant operators as
\begin{align}\label{decoderaction}
\Dcal_{n,j}(\ket{A_k^{m}}\bra{A_k^{m}})&=
\ket k\bra k,\\
\Dcal_{n,j}(\ket{B_{a,b}^{m}}\bra{A_b^{m}})&=\ket a\bra b,\\
\Dcal_{n,j}(\ket{A_a^{m}}\bra{B_{b,a}^{m}})&=\ket a\bra b,\\\Dcal_{n,j}(\ket{B_{a,k}^{m}}\bra{B_{b,k}^{m}})&=\ket a\bra b.
\end{align}
For $a=b$, we obtain
\begin{align}\label{appendixlogicaldiag}
    \Ncal_{n,j,t}(\ket{a}\bra{a})&=(t+rx(d-1))\ket{a}\bra{a}\nonumber\\&\quad +ry\sum_{\substack{k=0\\k\ne a}}^{d-1}\ket{k}\bra{k},
\end{align}
and for $a\ne b$,
\begin{align}\label{appendixlogicaloffdiag}
    \Ncal_{n,j,t}(\ket{a}\bra{b})&=\left[2\sqrt{rtx}+rx(d-2)\right]\ket{a}\bra{b}.
\end{align}
Substituting the values for $x,y$ and $r$ recovers Proposition~\ref{prop:logicalchannel}.
\newpage
\section{Overlap of maximally entangled projectors}\label{app:overlap-lemma}
For the sake of completeness, we include the proof for the following standard result~\cite{Zhang_2006}.
\begin{lemma}\label{lem:proj}
    Let $\Pi_S:=\Phi_{RD_S}\otimes I$ and $\Pi_T:=\Phi_{RD_T}\otimes I$ with $S\cap T=\emptyset$. Then $\Pi_S\Pi_T\Pi_S=\frac{1}{d^2}\Pi_S$ and $\norm{\Pi_S\Pi_T}_\infty=\frac{1}{d}$.
\end{lemma}
\begin{proof}
    Using the ordering $R\otimes D_S\otimes D_T$ and omitting the unaffected remaining systems, the projectors can be expanded as
    \begin{align}
        \Pi_S&=\frac{1}{d}\sum_{i,j,k=0}^{d-1}\ket{i,i,k}\bra{j,j,k},\\
        \Pi_T&=\frac{1}{d}\sum_{a,b,c=0}^{d-1}\ket{a,c,a}\bra{b,c,b}.
    \end{align}
    Composing once yields
    \begin{align}
        \Pi_T\Pi_S&=\frac{1}{d^2}\sum_{i,j,k,a,b,c}\ket{a,c,a}\bra{b,c,b}\ket{i,i,k}\bra{j,j,k}\nonumber\\
        &=\frac{1}{d^2}\sum_{a,j,c}\ket{a,c,a}\bra{j,j,c}\nonumber.
    \end{align}
    Composing again gives
    \begin{align}
        \Pi_S\Pi_T\Pi_S&=\frac{1}{d^3}\sum_{\ell,m,n,a,c,j}\ket{\ell,\ell,n}\bra{m,m,n}\ket{a,c,a}\bra{j,j,c}\nonumber
        \\
        &=\frac{1}{d^3}\sum_{\ell,j,n}\ket{\ell,\ell,n}\bra{j,j,n}\nonumber\\
        &=\frac{1}{d^2}\Pi_S\nonumber.
    \end{align}
Note that, since both $\Pi_S$ and $\Pi_T$ are orthogonal projectors, $\Pi_S\Pi_T\Pi_S = \Pi_S\Pi_T\Pi_T\Pi_S = (\Pi_S\Pi_T)(\Pi_S\Pi_T)^\adj$. Hence $\norm{\Pi_S\Pi_T}_{\infty}^2=\norm{\Pi_S\Pi_T\Pi_S}_\infty
    =\frac{1}{d^2}\norm{\Pi_S}_\infty=\frac{1}{d^2}$
Taking the square root proves the claim.
\end{proof}


\bibliography{apssamp}

\end{document}